\documentclass{aa}
\usepackage{graphics}
\begin{document}
    \thesaurus{1;  
              (05.05.1; 
               08.02.2; 
               08.09.2 CM Dra; 
		08.12.2; 
               08.16.2  
                   )}    
   \title{A search for Jovian-mass planets around CM Draconis using eclipse 
minima timing}

  \subtitle{}
 
   \author{
H.J. Deeg\inst{1}$^,$\inst{2} 
\and L.R. Doyle\inst{3}
\and V.P. Kozhevnikov\inst{4}
\and J.E. Blue\inst{5}
\and E. L. Mart\'{\i}n\inst{6}
\and J. Schneider\inst{7} }
	
   \offprints{H.J. Deeg}

   \institute{
Instituto de Astrof\'{\i}sica de Andaluc\'{\i}a, Granada, Spain (hdeeg@bigfoot.com) 
\and 
Instituto de Astrof\'{\i}sica de Canarias, E-38200 La Laguna (Tenerife), 
Spain
\and SETI Institute, 2035 Landings Drive, Mountain View, CA 94043, USA (ldoyle@seti.org)
\and Astronomical Observatory, Ural State University, Lenin ave. 51, Ekaterinburg 620083, Russia (valerij.kozhevnikov@usu.ru)
\and SRI International, 333 Ravenswood Ave., Menlo Park, CA 94025, USA 
\and Division of Geology and Planetary Sciences, Caltech MC 150-21, Pasadena, 
CA 91125, USA
\and CNRS-Observatoire de Paris, 92195 Meudon, France}
 
   \date{Received ; accepted }
 
   \maketitle
 
   \begin{abstract}

For the eclipsing binary system CM Draconis, eclipse minimum times
have been monitored with high precision between 1994 and
1999. Periodic deviations of minimum times from a linear ephemeris may
indicate the presence of an orbiting third body. Individual
measurements of 41 eclipse minimum times result in a standard
deviation from linear ephemeris of 5.74 seconds. A power spectral
analysis of the residuals reveals only one periodicity with more then
2 seconds amplitude. This feature, with a periodicity between 750 and
1050 days has an amplitude of 2.8$\pm$0.5 seconds, and is also present
with similar phases if the power spectral analysis is performed
independently for primary and secondary eclipses. It would be
compatible with a circumbinary planet of 1.5 -3 Jupiter masses at an orbital
distance of 1.1-1.45 AU to the binary barycenter. The assignation of a
planet to the CM Dra system can however only be upheld if this
periodicity can be followed in future observations for several
years. For low-mass eclipsing binary stars, the method of eclipse
minimum timing allows one to reach mass limits for the detection of
third bodies well below that feasible by radial velocity measurements.

 	\end{abstract}
 
%

\begin{figure*}
\resizebox{\hsize}{!}{\includegraphics{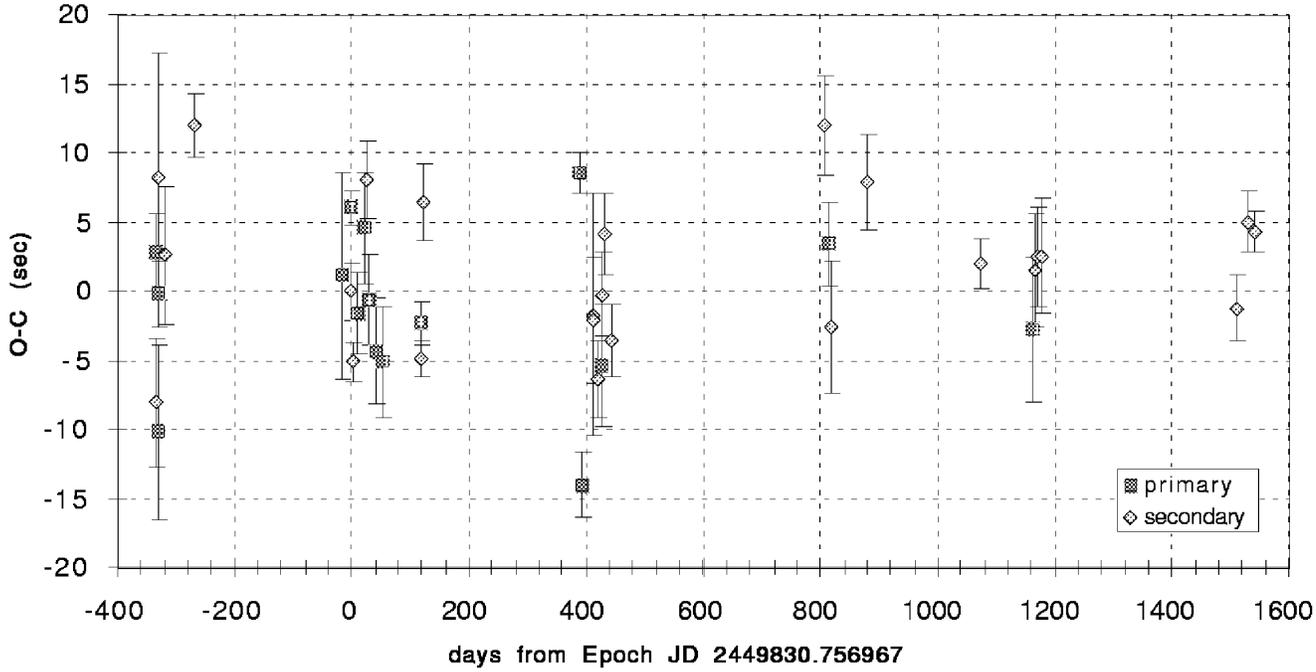}}
\caption{O-C residuals of CM Dra from 1994 to 1999. Six groups, corresponding 
to the yearly observing seasons (April-August), can be discerned}
\label{fig:figoc}
\end{figure*}

\section{Introduction}

It has long been known that the presence of a third body orbiting both
components of an eclipsing binary system will offset the binary from a
common binary/third-mass barycenter thereby causing a periodic shift
in the observed times of the binary eclipses. The amplitude of this
shift is given by \[ \delta T = M_\mathrm{P} a_\| / M_\mathrm{B} c ,\]
where $M_\mathrm{P}$ is the third body's mass, $a_\| = a \sin i$ the
third body's semi-major axis along the line of sight, $M_\mathrm{B}$
the mass of the binary system, and $c$ the speed of light. As pointed
out by \cite{schn95} and \cite{doyl98}, eclipse timings with a
precision of a few seconds could detect the presence of an orbiting
Jovian-mass object around a low-mass eclipsing binary system. Also,
\cite{doyl98} gives a sample of 250 eclipsing binaries for which
Jupiter-mass planets may be detectable by such studies. In this paper
we report on an analysis of eclipse timings that were obtained as part
of a photometric search for extrasolar planetary transits undertaken
during the six years 1994 - 1999 around the M4.5/M4.5 binary CM Dra by
the TEP project (\cite{obspap} --~further TEP1~--~; \cite{TEP2} --~further TEP2~--~)

\section{Data and Analysis}

Eclipse minimum times were obtained from photometric time series data
of CM Dra; the photometric reduction pipeline is described in TEP1.
Photometric data included in this analysis have a maximum photometric
rms error of 0.7\% and were corrected for nightly extinction
variations.  Eclipses of CM Dra were extracted from the data with a
cut-off of $\Delta m > 0.1 \mathrm{mag}$ from the off-eclipse baseline.  The
eclipse minimum times were then measured with a 7-segment
Kwee-van-Woerden algorithm (\cite{kvw}), and converted to heliocentric
Julian dates with the `setjd' routine in IRAF. The entire lightcurve
with 1014 hours of coverage, taken by all telescopes of the TEP
project between 1994 and 1999, contains 81 eclipses for which O-C
times were measurable.  For further analysis, however, we selected
only data from a subset of telescopes which delivered the most
consistent results for timing.  These were the Crossley telescope at
Lick Observatory, the JKT, INT and IAC80 telescopes at the Instituto
de Astrof\'{\i}sica de Canarias, the 0.6m at Kourovka Observatory and
the 1.2m of the Observatoire de Haute Provence.  Inconsistent minimum
times from those telescopes whose data were rejected were most likely
caused by imprecise recording of the time, which depends in most
systems on the computer that archives the data. We also excluded the
eclipses observed by \cite{lacy77}, whose two primary eclipses had
discrepancies of 25 seconds between them (based on a re-analysis of
Lacy's data, see TEP 1). From the remaining data only those minimum
times were kept where the lightcurve covered the entire ingress and
egress of each eclipse without significant 'holes', and where the
formal error given by the Kwee-van-Woerden algorithm was less than 10
seconds.  The resulting sample, to be further investigated, contains
minima timings of 16 primary and 25 secondary eclipses.

\section{Results}

In Fig.~\ref{fig:figoc}, the O-C (observed - computed) minimum times
of these 41 eclipses are plotted. The computed minimum times $T_n$ are
based on the linear ephemeris given by TEP1, where $T_{n} =
T_\mathrm{o} + nP_\mathrm{orb}$, with a period of $P_\mathrm{orb}=
1.268\,389\,861 \pm 0.000\,000\,005 $days, an epoch of primary
eclipses of $T_\mathrm{o} = \mathrm{HJD} 2449830.757\,00 \pm
0.000\,01$, and an epoch of secondary eclipses of $ T_\mathrm{o} =
\mathrm{HJD} 2449831.390\,03 \pm 0.000\,01$. This ephemeris was
derived from eclipses measured between 1994 and 1996. The eclipses
observed afterwards, in 1997-1999, do not exhibit any trends away from
that ephemeris, and therefore no attempts have been made to derive a
new one. The standard deviation of all O-C times from 1994-1999
against the elements listed above is 5.87 seconds for primary, 5.47
s for secondary, and 5.74 s for both eclipses together.

To evaluate the minimum times for the presence of periodicities, we
performed a power spectral analysis using the method of sine-wave
fitting common in solar oscillation studies (\cite{kjel92}). In this
method, sine waves with increasing periods are fitted to the O-C
values, using amplitude and phase as fitting parameters. (This is
identical to fitting a sinusoidal ephemeris, $T_n = T_\mathrm{o} + n
P_\mathrm{orb} + A \sin[2 \pi (t-\tau)/P + \kappa]$ with stepwise
increasing periods $P$, and recording amplitude $A$ and phase $\kappa$
of the best fit.) This method has the advantage over the Lomb
periodogram spectral analysis (\cite{lomb76}, \cite{press92}) that
amplitudes are derived with an absolute scale, whereas the Lomb method
derives only relative amplitudes.

\begin{figure}
\resizebox{\hsize}{!}{\includegraphics{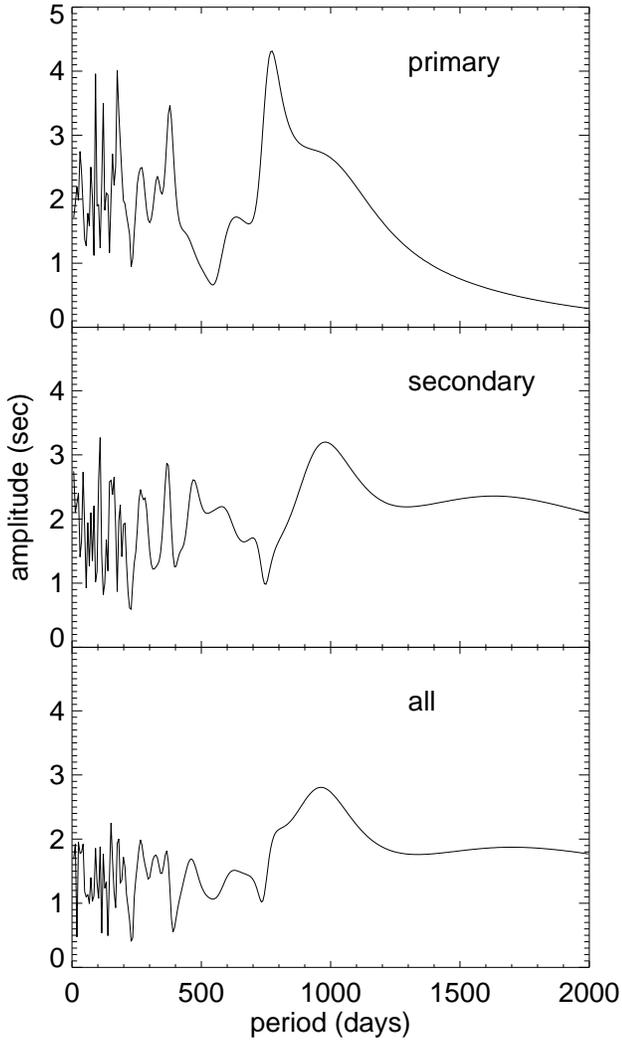}}
\caption{O-C power spectra,  from primary (upper panel), secondary (central panel), and from all eclipse times (lower panel)}
\label{fig:figpow}
\end{figure}

\begin{figure}
\resizebox{\hsize}{!}{\includegraphics{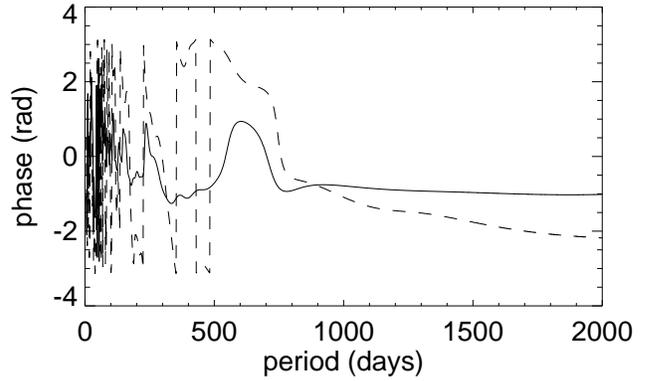}}
\caption{Phases of O-C power spectra. Solid line is from primary eclipses, and dashed line from secondary ones. The phases are wrapped to the range $\pm\pi$.}
\label{fig:figfase}
\end{figure}

The power spectra (Fig.~\ref{fig:figpow}) were obtained separately for
primary and secondary eclipses, as well as for both eclipses combined.
As can be seen, the highest peaks have amplitudes around 4
seconds. The only notable feature is a peak between 750 and 1050 days,
occurring in both kinds of eclipses. This is the only feature
significantly above 2 seconds amplitude in the power diagram of
primary and secondary eclipses combined (Fig 2c). It has a maximum
amplitude of 2.8 seconds at a period of 970 days. Also, the phases of
the powerspectra (Fig.~\ref{fig:figfase}) are close for primary and
secondary eclipses in that period range, being identical at a period of
890 days. Above periodicities of 2000 days, there is a smooth decay of
spectral power, which is a consequence of the length of coverage of
our data - the distance between the first and the last eclipse in the
data set is 1879 days.

\section{Discussion}
The power spectra (Fig.\,2) indicate that there are no periodic O-C
minimum time variations with amplitudes of larger then 3-4 seconds
present, for all periods less than 2000 days.  This absence of
amplitude variations allows us to \emph{exclude} the presence of very
massive planets around the CM Dra system, as indicated by the hatched
region in the search-space diagram (Fig.\,3).  Excluding the peak around
1000 days period, the power spectra from all data (lower panel of Fig.\,2) is
relatively flat with an amplitude of about 2 seconds. This white-noise
like flatness indicates an intrinsic imprecision in our data of about
2 seconds.  This is most likely the results of the precision of the
eclipse minima times being limited by the photometric noise of the
eclipse lightcurves. O-C deviations of about 2 seconds constitute
therefore a lower detection limit.  Finally, the peak in the power
spectra between 750 and 1100 days with an amplitude of $2.5 \pm 0.5 $
seconds and a good match of phases from primary and secondary eclipses
may be the consequence of a third body, but is close to the
observational noise. If this amplitude variation is caused by a third
body, it would correspond to a circumbinary planet of 1.5-3 Jupiter masses at an
orbital distance from CM Dra of 1.1 - 1.45 AU. We note that such a
body would cause a periodic variation in the radial velocity of CM Dra
with an amplitude of $65 \pm 20 \mathrm{m s}^{-1}$. Though sufficient precision
to detect such radial velocities amplitudes has routinely been
obtained in planetary detection programs, these program are always
concerned with single stars. For eclipsing binaries, the mutual
orbiting of the binary components causes large radial velocity
amplitudes on the order of km/s, which obstruct the separation of the
much smaller radial velocity amplitudes from a third body.  In the
case of CM Dra, the velocities of the binary components reported by
\cite{metcalfe96} are 72 and 78 km/s, and the precision of these data
would only allow the separation of third body amplitudes of more than
200 m/s (\cite{latham}).  Finally, the limited time-baseline of our
observations does not allow the detection of periodicities longer then
about 2000 days. The absence of very heavy third bodies with periods
up to a few times longer is however rather certain due to the good
general adherence of the O-C times to a linear ephemeris.  Influences
from third bodies {\emph within} the Solar System onto the
heliocentric eclipse minimum times are not of consequence. The
strongest influence, by Jupiter, causes a 12-yearly deviation, but due
to the high ecliptic latitude of CM Dra ( $76.3 \deg$) its amplitude
is limited to 0.58 seconds.

The \emph{absence} of periodicities above 3-4 seconds amplitude - 
and the exclusion of corresponding massive planets - may be stated with 
certainty, even if the data analysis may not have accounted for every factor 
that may introduce spurious periodicities.  The claim by \cite{guin98} of 
a periodicity in minimum times of 70 days with an amplitude of 18 seconds, 
corresponding to a third body with a mass of $0.01 M_\odot$, is clearly 
invalidated (see also \cite{iauc6875}).

The \emph{presence} of apparent periodicities with $\approx $ 3
seconds may however also be a consequence of slowly changing starspots
which distort the symmetry of the eclipses.  Although we have not been
able to find any relevant variations in lightcurves of CM Dra through
the different observing seasons 1994-1999, the possibility of
starspots can not be entirely excluded. In any case, further
monitoring with high precision minimum timing of the CM Dra system is
needed to ascertain the continuing presence of the 700-1050 day
periodicity.

\begin{figure}
\resizebox{\hsize}{!}{\includegraphics{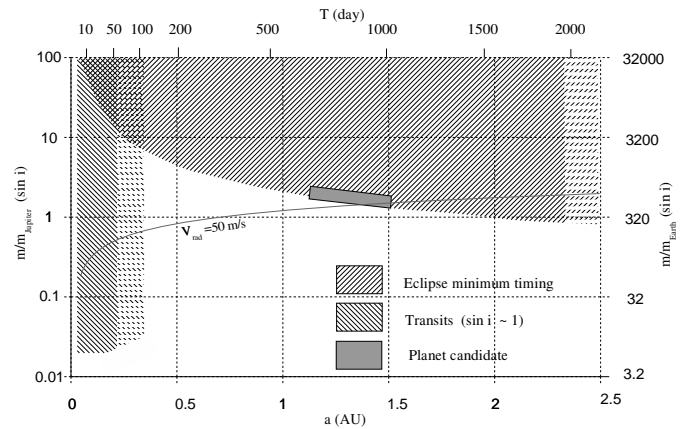}}
\caption{Search space covered by the O-C timing observations and
by transit observations reported in TEP1, TEP2.  The hatched regions
are those where planets can be excluded from the eclipse minimum
timing and the transit search reported in TEP2.  The small rectangular
gray region corresponds to the planet candidate from the power-spectra
based on O-C minimum timing. Regions left blank are those where these
detection methods have not had sufficient sensitivity. Also indicated
is a line where an orbiting third body would cause a radial velocity
variation of 50 m/s in the spectrum of the eclipsing binary.}
\label{fig:figdiscover}
\end{figure}

Fig.~\ref{fig:figdiscover} shows the search space of exoplanets
around CM Dra covered by the eclipse timing observations described
here and by the observations of transits from TEP1 and TEP2. The
transit observations covered coplanar planets ($\sin i \approx 1$) on
short period orbits, between 7 days (the shortest stable orbit around
CM Dra) and 60 days (as a limit where observational coverage gets
sparse), with a maximum detectable periodicity of 100 days (the limit
where even coplanar planets would not cause transits because of the
89.82$\degr$ inclination of the system). We assumed a mass limit of
$m/m_\mathrm{Earth} \approx 10$, corresponding to the lower size limit of
about 2.5 Earth Radii for detections with transits. The lower mass
limit from the O-C timing method is derived from the absence of
amplitudes over 2.5 seconds, except between 700 and 1050 days, were a
planet candidate is indicated. 

The two methods employed do cover rather complementary regimes:
Whereas the strength of the transit method is the detection of
relatively small planets on close orbits, O-C minimum timing is best
for the detection of long period planets with at least Jupiter-like
masses. The usefulness of the radial velocity method is limited in
binary systems, though it might also lead to the discovery of massive
third bodies around them. To verify the persistence of the 700-1050 day
periodicity and the possibility of a planet, observations of CM Dra's
eclipse minimum times need to be continued during the the next several
years.

\section {Acknowledgments}
We thank R. Garrido, A. Gimenez and the anonymous referee for
helpful suggestions and Ayvur Akalin for help with the Kwee-Van-Woerden algorithm.

\end{document}